# Polarizability of ultracold Rb$_2$ molecules in the rovibrational ground state of a$^3\Sigma_u^+$

**Markus Deiß, Björn Drews, and Johannes Hecker Denschlag**

Institut für Quantenmaterie and Center for Integrated Quantum Science and Technology IQ$^{ST}$, Universität Ulm, 89069 Ulm, Germany

**Nadia Bouloufa-Maafa, Romain Vexiau, and Olivier Dulieu**

Laboratoire Aimé Cotton, CNRS, Université Paris-Sud, ENS Cachan, Bât. 505, Campus d'Orsay, 91405 Orsay Cedex, France

E-mail: johannes.denschlag@uni-ulm.de

**Abstract.** We study, both theoretically and experimentally, the dynamical polarizability $\alpha(\omega)$ of Rb$_2$ molecules in the rovibrational ground state of a$^3\Sigma_u^+$. Taking all relevant excited molecular bound states into account, we compute the complex-valued polarizability $\alpha(\omega)$ for wave numbers up to 20000 cm$^{-1}$. Our calculations are compared to experimental results at 1064.5 nm ($\sim 9400$ cm$^{-1}$) as well as at 830.4 nm ($\sim 12000$ cm$^{-1}$). Here, we discuss the measurements at 1064.5 nm. The ultracold Rb$_2$ molecules are trapped in the lowest Bloch band of a 3D optical lattice. Their polarizability is determined by lattice modulation spectroscopy which measures the potential depth for a given light intensity. Moreover, we investigate the decay of molecules in the optical lattice, where lifetimes of more than 2 s are observed. In addition, the dynamical polarizability for the X$^1\Sigma_g^+$ state is calculated. We provide simple analytical expressions that reproduce the numerical results for $\alpha(\omega)$ for all vibrational levels of a$^3\Sigma_u^+$ as well as X$^1\Sigma_g^+$. Precise knowledge of the molecular polarizability is essential for designing experiments with ultracold molecules as lifetimes and lattice depths are key parameters. Specifically the wavelength at $\sim 1064$ nm is of interest, since here, ultrastable high power lasers are available.

## 1. Introduction

Owing to the extraordinary control over the internal and external degrees of freedom, ultracold molecules trapped in an optical lattice represent a system with many prospects for studies in ultracold physics and chemistry [1, 2], the realization of molecular condensates [3], precision measurements of fundamental constants [4, 5, 6, 7] and quantum computation [8, 9] and simulation [10]. In the recent years, several groups have realized the preparation of optically trapped vibrational ground state ($v = 0$) molecules in either the lowest lying singlet or triplet potential [11, 12, 13, 14, 15]. Experiments with these molecules, e.g. ultracold collisions, are typically carried out in optical lattices or optical dipole traps [16, 17, 18]. In these environments, precise knowledge of the dynamical polarizability of molecules is important for well controlled experiments.

The $Rb_2$ molecule is one of the few ultracold molecular species currently available, with which benchmark experiments for nonpolar molecules can be carried out. Here, we investigate the dynamical polarizability $\alpha(\omega)$ of a $Rb_2$ triplet molecule in the lowest rovibrational level of $a^3\Sigma_u^+$. A similar analysis for $Cs_2$ regarding the electronic ground state $X^1\Sigma_g^+$ was previously carried out by Vexiau *et al.* [19]. In addition to calculations of the frequency dependent dynamical polarizability $\alpha(\omega)$, we present measurements of the real part $\text{Re}\{\alpha(\omega)\}$ at a wavelength of $\lambda = 1064.5\,\text{nm}$. The experiments are performed with molecules trapped in the lowest Bloch band of a cubic 3D optical lattice which consists of three standing light waves with polarizations orthogonal to each other. By carrying out modulation spectroscopy on one of the standing light waves, we map out the energy band-structure for various light intensities. From these measurements $\text{Re}\{\alpha(\omega)\}$ is determined. Our experimental findings at $\lambda = 1064.5\,\text{nm}$ and also those at 830.4 nm [13] agree well with the calculations. In addition, we experimentally investigate the decay time of the deeply bound molecules in a 3D optical lattice at 1064.5 nm for various lattice depths. Here, lifetimes of more than 2 s are observed. Furthermore, we present numerical results for the dynamical polarizabilities of the rovibronic ground state, i.e., the lowest rovibrational level of the $X^1\Sigma_g^+$ potential. For convenient application of our results, we provide a simple analytical expression and the corresponding effective parameters which can be used to reproduce the dynamical polarizabilities for all vibrational levels of both, the lowest singlet as well as triplet state outside the resonant wavelength regions.

This article is organized as follows. In section 2, we give a brief, general introduction to the dynamical polarizability $\alpha(\omega)$ of a homonuclear diatomic molecule. Sections 3 to 6 describe the calculations and measurements related to the polarizability of $^{87}Rb_2$ in the rovibrational ground state of $a^3\Sigma_u^+$, along with a comparison of our results to reference values from literature. In section 7 we present calculations of the polarizability for the rovibrational ground state of $X^1\Sigma_g^+$. Afterwards, section 8 provides a simple expression which parametrizes the polarizability for all vibrational levels of the lowest singlet as well as triplet state. Tables with the corresponding parameters can be found in the Supplemental Material [20].

## 2. Interaction of a diatomic molecule with light

When a nonpolar molecule is subject to a linearly polarized electric field $\vec{E} = \hat{\varepsilon} E_0 \cos(\omega t)$ with amplitude $E_0$ and unit polarization vector $\hat{\varepsilon}$, a dipole moment $\vec{p} = \overset{\leftrightarrow}{\alpha}(\omega)\vec{E}$ is induced. In general, $\overset{\leftrightarrow}{\alpha}(\omega)$ is a tensor (see, e.g., [21]). For the sake of simplicity, we restrict ourselves to Hund's case (b) molecules in the lowest rotational level $N = 0$ of the nuclei, for which only the scalar isotropic polarizability $\alpha(\omega)$ is relevant (see, e.g., [21, 13]). Here, $\vec{N} = \vec{L} + \vec{R}$, where $\vec{L}$ denotes the total electronic orbital angular momentum and $\vec{R}$ is the mechanical rotation of the atomic pair. The complex dynamical polarizability characterizes the response of a molecule to the electric field expressed by photon scattering and the ac Stark shift of molecular levels. This shift is directly linked to the dipole interaction potential

$$U = -\langle \vec{p} \cdot \vec{E} \rangle / 2 = -\frac{1}{4} \mathrm{Re}\{\alpha(\omega)\} E_0^2 \tag{1}$$

and therefore to the real part of the polarizability $\mathrm{Re}\{\alpha(\omega)\}$ (see, e.g., [22]). In Eq. (1), the angled brackets $\langle \ldots \rangle$ indicate time averaging. We note, that the dipole potential is attractive, when the sign of $\mathrm{Re}\{\alpha(\omega)\}$ is positive and repulsive otherwise. The imaginary part of the polarizability $\mathrm{Im}\{\alpha(\omega)\}$ is related to the power $P_{\mathrm{abs}}$ absorbed by the oscillator from the driving field, since

$$P_{\mathrm{abs}} = \langle \dot{\vec{p}} \cdot \vec{E} \rangle = \frac{1}{2} \omega \, \mathrm{Im}\{\alpha(\omega)\} E_0^2. \tag{2}$$

We calculate the dynamical polarizability $\alpha(\omega)$ following the method described in Ref. [19]. The generic expression of the polarizability for a diatomic molecule in a state $|i\rangle$ is

$$\alpha(\omega) = \frac{2}{\hbar} \sum_f \frac{\omega_{if} - i\frac{\gamma_f}{2}}{\left(\omega_{if} - i\frac{\gamma_f}{2}\right)^2 - \omega^2} \left| \langle f|\vec{d} \cdot \hat{\varepsilon}|i\rangle \right|^2. \tag{3}$$

Here, the angled brackets refer to the spatial integration over all internal coordinates of the system. The summation covers all the accessible dipole transitions with frequency $\omega_{if}$ and transition electric dipole moment $\vec{d}$ from the initial state $|i\rangle$ to final states $|f\rangle$ with line width $\gamma_f$.

## 3. Calculation of $\alpha(\omega)$ for $a^3\Sigma_u^+$ molecules

### 3.1. Relevant transitions

If the molecule is initially in a vibrational level $v_a$ of the $a^3\Sigma_u^+$ state, all rovibrational levels (including the continuum) of the electronic potentials with $^3\Sigma_g^+$ and $^3\Pi_g$ symmetry need to be accounted for in Eq. (3). As $N = 0$ in the initial $a^3\Sigma_u^+$ state, only transitions towards final levels with total angular momentum $J = 0, 1, 2$ must be considered. Here, $\vec{J} = \vec{S} + \vec{N}$, where $\vec{S}$ is the total electronic spin. Therefore, when considering a diatomic molecule it is usual to define two contributions to the isotropic polarizability $\alpha$: the parallel polarizability $\alpha_\parallel$ along the molecular axis $\hat{Z}$, which is related to $d_Z$, and the perpendicular polarizability $\alpha_\perp$, which is related to $d_X = d_Y$. In general, $\alpha_\parallel$ involves $^3\Sigma^+ \to {}^3\Sigma^+$ transitions and $\alpha_\perp$ is related to $^3\Sigma^+ \to {}^3\Pi$ transitions. One can show that $\alpha = (\alpha_\parallel + 2\alpha_\perp)/3$ (see, e.g., Refs. [19, 21]).

The expression given by Eq. (3) deals only with the transitions involving the two valence electrons of $Rb_2$. Following Ref. [23], the contribution to the polarizability of the two $Rb^+$ cores, hereafter referred to as $\alpha_c$, must be taken into account, and is added to the results of Eq. (3). More details about this quantity are discussed in Appendix A.1.

The first step of the calculations is to collect a set of accurate molecular potential energy curves (PECs) and transition electric dipole moments (TEDMs). The $a^3\Sigma_u^+$ PEC is taken from the spectroscopic study of Ref. [24]. For the excited molecular states and the related TEDMs from the $a^3\Sigma_u^+$ state, we use the same data as Refs. [25, 26], which we report in the Supplemental Material [20] attached to the present paper, for convenience. The PECs are displayed in Fig. 1(a), while the TEDMs are drawn in Fig. 10 (see Appendix A.2). These data are obtained by the quantum chemistry approach described in details in Ref. [27]. Briefly the $Rb_2$ molecule is considered as two valence electrons moving in the field of the two ionic $Rb^+$ cores, which are represented by a large effective core potential (ECP) including a core polarization potential (CPP) [28, 29]. A full configuration interaction (FCI) is then performed on the two valence electrons, using a large gaussian basis set [30], with the CIPSI quantum chemistry code developed at Université Paul Sabatier in Toulouse. It is worth mentioning that partial spectroscopic information is available on the $1^3\Sigma_g^+$ state [31], the $1^3\Pi_g$ state [32], and on the $2^3\Pi_g$ state [24], but no complete PEC has been extracted in these studies. As discussed for instance in Refs. [25, 32], the computed PECs are suitable to reproduce the observed data provided that they are slightly shifted in frequency (in terms of $\omega/(2\pi c)$, by at most $100\,cm^{-1}$). We will estimate in section 5 the limited influence of such shifts on the results reported in the present work. Finally, the vibrational wave functions of levels $|i\rangle$ and $|f\rangle$ for the summation are obtained using the Mapped Fourier Grid Hamiltonian representation [33, 34].

## 3.2. Results

The real and imaginary parts of the dynamical polarizability $\alpha_{v_a=0}(\omega)$ of a molecule in the vibrational ground state of the $a^3\Sigma_u^+$ potential are displayed in Fig. 1(b) and (c) as functions of the trapping laser frequency. The polarizabilities are expressed in atomic units (a.u.), which can be converted into SI units according to $1\,a.u. = 4\pi\varepsilon_0 a_0^3 = 1.649 \times 10^{-41}\,Jm^2V^{-2}$, where $a_0$ denotes the Bohr radius and $\varepsilon_0$ is the vacuum permittivity. Note, for some applications, e.g., considerations related to the ac Stark shift, units of $HzW^{-1}cm^2$ (1 a.u. corresponds to $4.6883572 \times 10^2\,HzW^{-1}cm^2$) are advantageous. The sum in Eq. (3) has been truncated to include only the vibrational levels of the four lowest $^3\Sigma_g^+$ states and the three lowest $^3\Pi_g$ states. Furthermore, electric-dipole-forbidden transitions are not considered in the sum, as they would appear as very weak and narrow resonances in the polarizability. The associated molecular data are collected in the Supplemental Material [20]. For simplicity, the natural lifetime $\tau_f = (\gamma_f)^{-1}$ has been fixed to $10\,ns$ ($\gamma_f \approx 2\pi \times 15\,MHz$) for all the excited molecular levels.

Strongly oscillating patterns in both $Re\{\alpha_{v_a=0}(\omega)\}$ and $Im\{\alpha_{v_a=0}(\omega)\}$ [see Figs. 1(b) and (c), respectively] correspond to frequency ranges of strong absorption which should be disregarded for trapping purpose. The real part smoothly increases from the static polarizability $\alpha_{v_a=0}(\omega = 0) = 698.5$ a.u. up to the bottom of the $1^3\Sigma_g^+$ potential well, reaching 3147 a.u. at the wavelength of the trapping laser used in the present experiment (1064.5 nm). In the same region the imaginary part increases from about $10^{-5}$ a.u. at $\omega = 0$ to $10^{-3}$ a.u. at the trapping laser frequency which leads to a correspondingly larger photon scattering rate.

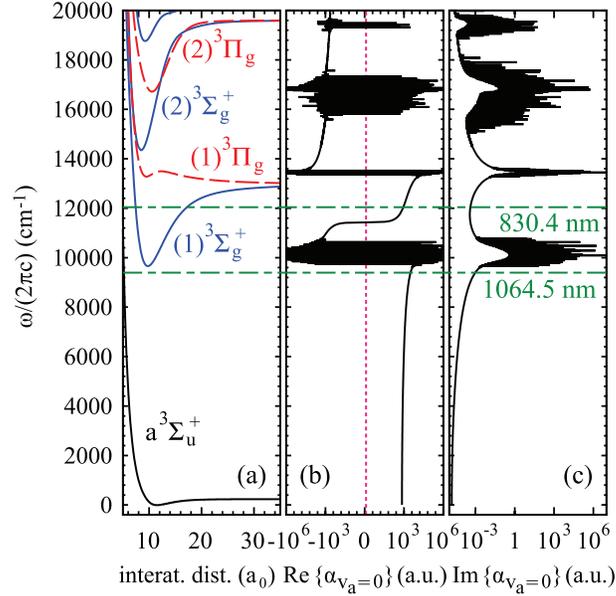

**Figure 1.** (a) $^3\Sigma_g^+$ (solid blue lines) and $^3\Pi_g$ (dashed red lines) potential curves of $Rb_2$ [25, 26]. The $a^3\Sigma_u^+$ potential is drawn in black. In (b) and (c) the real and imaginary parts of the dynamical polarizability $\alpha_{v_a=0}$ of $Rb_2$ molecules in the rovibrational ground level ($v_a = 0$, $R = 0$) of the $a^3\Sigma_u^+$ molecular state are shown as a function of $\omega/(2\pi c)$. The two wavelengths used in the experiments are indicated by dashed horizontal lines. Furthermore, the red dashed vertical line in (b) represents zero polarizability.

It is difficult to provide a well-defined error bar on the theoretical values of the dynamical polarizabilities as their accuracy depends on the considered wavelength. Various causes of global inaccuracies have been analyzed in depth in Ref. [35]. First, the choice of a constant radiative lifetime for all excited levels in Eq. (3) influences only the strongly oscillating regions of the polarizability, changing the amplitude of the resonances. We checked that this approximation has no effect in the smoothly varying regions which are relevant for trapping experiments. We verified also that adding a couple of upper electronic states in the sum of Eq. (3) contributes to the polarizability for less than 1%. Moreover, the first excited $\Sigma$ and $\Pi$ states contribute together for more than 90% to the polarizability. Usually, these are the most well known states either because accurate spectroscopic results are available, or because they are well-determined by quantum chemistry calculations. The accuracy of TEDMs is tedious to analyze as their experimental determination relies on line intensities which are difficult to measure accurately. However, one argument in favor of the accuracy of the TEDMs results when comparing the values obtained from different methods. For instance, in Refs. [36, 37], with respect to various alkali-metal dimers, the TEDMs computed by two different methods are found to agree within 2%. Finally, an indication for the accuracy of the present work is provided by the measurement of the dynamical polarizability for the $v = 0$, $J = 0$ level of the $Cs_2$ electronic ground state at 1064.5 nm, which is quite far away from the lowest resonant region [19]. The experimentally determined value with respect to $\lambda = 1064.5$ nm is $2.42(15) \times \alpha_{Cs}$ [39], where $\alpha_{Cs}$ is the dynamical polarizability of the Cs atom. This is in remarkable agreement with the computed value of $2.48 \times \alpha_{Cs}$ [19].

## 4. Measurement of $Re\{\alpha(\omega)\}$ for $a^3\Sigma_u^+$

### 4.1. Experimental setup and measurement scheme

The experiments presented in this work are carried out with a pure sample of about $1.5 \times 10^4$ $^{87}Rb_2$ molecules prepared in the rovibrational ground state of the $a^3\Sigma_u^+$ potential and trapped in a 3D optical lattice. There is no more than a single molecule per lattice site and the temperature of the sample is about 1 $\mu$K. As described in detail in Refs. [13, 21, 38], the molecules are prepared as follows. An ultracold thermal cloud of spin-polarized $^{87}Rb$ atoms ($f_{Rb} = 1$, $m_{f,Rb} = 1$) is adiabatically loaded into the lowest Bloch band of a 3D optical lattice at a wavelength of $\lambda = 1064.5$ nm. The lattice is formed by a superposition of three linearly polarized standing light waves with polarizations orthogonal to each other, see Fig. 2(a). The three lattice beams are derived from the same laser source with a linewidth of a few kHz and have relative intensity fluctuations of less than $10^{-3}$. In order to avoid interference effects, the frequencies of the standing waves are offset by about 100 MHz relative to each other. At the location of the atomic sample the beam waists ($1/e^2$ radii) are about 130 $\mu$m and the maximum available power per beam is about 3.5 W. By slowly crossing the magnetic Feshbach resonance at 1007.4 G we produce weakly bound diatomic molecules. After a purification step which removes remaining atoms, a STIRAP (stimulated Raman adiabatic passage) is performed at 1000 G, transferring the dimers into the rovibrational ground state ($v_a = 0$, $N = 0$, $m_N = 0$) of the $a^3\Sigma_u^+$ potential. Here, $m_N$ is the projection of $N$ on the quantization axis defined by the direction of the magnetic field $\vec{B}$ [cf. Fig. 2(a)]. The molecule has positive total parity, total electronic spin $S = 1$, total nuclear spin $I = 3$ and is further characterized by the quantum number $f = 2$ ($\vec{f} = \vec{S} + \vec{I}$). Moreover, the total angular momentum is given by $F = 2$ ($\vec{F} = \vec{f} + \vec{J}$) and its projection is $m_F = 2$. Henceforth, we simply refer to these molecules as "$v_a = 0$ molecules".

According to Eq. (1), $Re\{\alpha(\omega)\} = 4|U|/E_0^2$, i.e., the real part of the dynamical polarizability can be determined by measurements of the potential depth $|U|$ and the electric field amplitude $E_0$ of an optical trap. For the case of a cubic 3D optical lattice with orthogonal polarizations, the trapping potential is given by $V(x, y, z) = \sum_{\beta=x,y,z} V_\beta(\beta)$ where the

$$V_\beta(\beta) = -|U|_\beta \cos^2\left(k\beta + \phi_\beta\right) \tag{4}$$

represent the contributions of the standing waves of directions $\beta = x, y, z$ with $k = 2\pi/\lambda$ being the wave number of the lattice beams. We now only consider the part of the lattice in the vertical $z$ direction since this axis is the only one relevant for the measurements of the potential depth in the present work. Therefore, we define $V_z(z) \equiv V(z)$, $|U|_z \equiv |U|$ and $\phi_z \equiv \phi$. The phase $\phi$ is a function of the laser wavelength $\lambda$ because the standing light wave is created

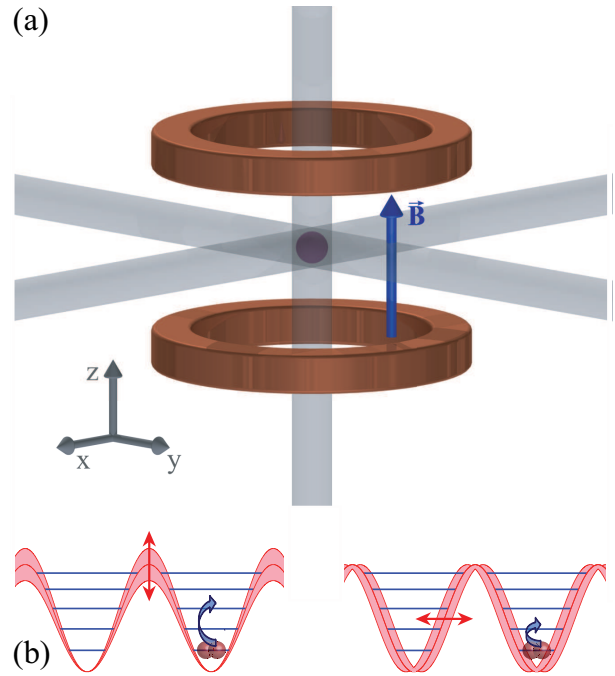

**Figure 2.** (a) Experimental scheme. Three retroreflected laser beams with polarizations orthogonal to each other form a 3D optical lattice. A trapped cloud of $Rb_2$ molecules is sketched in the intersection of the laser beams. The optical lattice is located between two Helmholtz coils which create a magnetic field $\vec{B}$. In our setup, the direction of $\vec{B}$ represents the quantization axis. (b) Illustration of amplitude (left) and phase (right) modulation spectroscopy.

by retroreflecting the laser beam from a fixed mirror at position $z_m$. It is given by $\phi = 4\pi z_m/\lambda$ at $z = 0$.

### 4.2. Lattice modulation spectroscopy

In order to obtain $|U|$, we carry out lattice modulation spectroscopy (see, e.g., [12, 40, 41, 42]). For this, we either modulate $|U|$ by periodically changing the intensity of the standing light wave (amplitude modulation) or we modulate the phase $\phi$ by periodically changing the laser wavelength $\lambda$ (phase modulation). Resonant amplitude (phase) modulation drives transitions from the lowest Bloch band ($n = 0$), in which the molecules have been initially prepared, to even (odd)-numbered excited lattice bands [see Fig. 2(b)]. This can cause either direct loss from the trap owing to heating (see, e.g., [40]) or molecules in higher lattice bands collide with each other and those of the lowest Bloch band, respectively, resulting in decay to nonobservable states. In consequence, resonant excitation leads to a decreased molecular signal in our measurements.

For amplitude modulation spectroscopy, we modulate the intensity of the lattice laser beam sinusoidally by a few percent. When performing phase modulation spectroscopy, we modulate the laser frequency by a few MHz corresponding to a phase difference on the order of a few $10^{-2}$ rad as $z_m \sim 0.4$ m. The modulation duration is typically on the order of 1 ms.

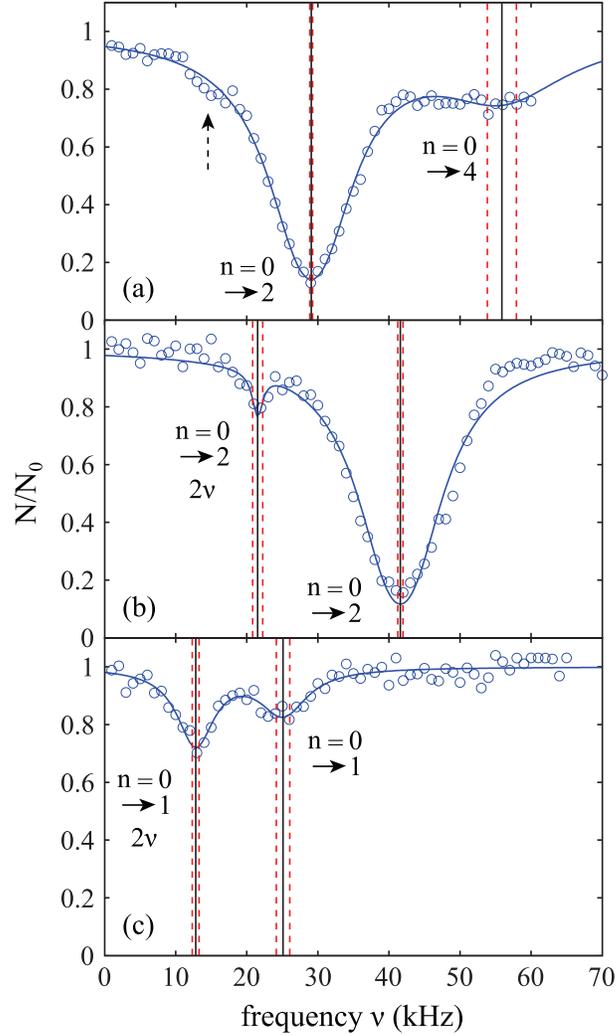

**Figure 3.** Amplitude [(a) and (b)] and phase (c) modulation spectra of weakly bound Feshbach molecules (a) and molecules in the rovibrational ground state of the a$^3\Sigma_u^+$ potential [(b) and (c)]. We measure the fraction of remaining molecules $N/N_0$ as a function of the modulation frequency $\nu$. The statistical error of each data point is in the range of $\pm(0.05-0.15)$. Here, the numbers $N_0$ are given by the asymptotic limits of Lorentzian fits (solid blue lines). The resulting center frequencies of the resonances are illustrated as black vertical lines, while the red dashed lines indicate the corresponding uncertainties. We note, that the mean intensity of the lattice beam used for modulation in (b) is 30% less than in (a) and (c).

At the end of each experimental cycle (which takes about 40 s), the remaining number $N$ of molecules is measured. We only find molecules in the lowest Bloch band, not in higher bands. In order to determine the molecule number, we reverse the STIRAP and dissociate the resulting Feshbach dimers by sweeping over the Feshbach resonance. Then, the generated atoms are detected via absorption imaging. By comparing the resonant transition frequencies observed in the modulation spectra to the energy band-structure of the sinusoidal lattice, the lattice depth $|U|$ is deduced. This will be explained in detail further below.

Figure 3 shows measured excitation spectra of Feshbach (a) and $\nu_a = 0$ (b,c) molecules, obtained via amplitude or phase modulation spectroscopy. A single data point typically consists of 5 to 30 repetitions of the experiment (For a given spectrum the number of repetitions is constant). Fig. 3(a) as well as (b) exhibit a prominent resonance after amplitude modulation. This resonance is related to a transition from the lowest Bloch band ($n = 0$) to the second excited lattice band ($n = 2$). Spectrum (a) for Feshbach molecules in addition shows a broad shoulder at around 50 kHz which we attribute to a resonant transition from $n = 0$ to $n = 4$. Due to the large width of this resonance the uncertainty in the determination of its center frequency is relatively large. Spectrum (b) in Fig. 3 also features a second resonance dip, but here it is located at about half the frequency of the prominent one. This resonance dip can be assigned to a transition from the lowest Bloch band to the second excited band, involving two identical "quanta" with frequency $\nu$. It is known (see, e.g., [40]) that such subharmonic resonances exist. To be consistent, a similar sub-harmonic resonance dip should be present in Fig. 3(a) at about 15 kHz (as indicated by the vertical, dashed arrow). Indeed, at that position the data points seem to be systematically below the fit curve with respect to the prominent peak. However, the corresponding signal (if at all) is very weak, partially due to its position at the steep flank of the prominent resonance.

Now, we turn to Fig. 3(c), which shows an excitation spectrum after phase modulation for $\nu_a = 0$ molecules. We observe two resonances of similar strength, both of which we assign to the transition from $n = 0$ to $n = 1$. The dip at lower frequency is again a subharmonic resonance. Surprisingly, it is stronger than the harmonic one at about 25 kHz. We attribute this to a purely technical issue, as the strength of the phase modulation varied with the frequency in our setup. However, we have verified the assignment of the resonances by comparison to the corresponding amplitude modulation spectra.

We calculate the Bloch bands by diagonalizing the Hamilton operator for the lattice in 1D (neglecting gravitation),

$$H = -\frac{\hbar^2}{2m} \frac{\partial^2}{\partial z^2} + V(z),$$

(5)

which is particularly simple in momentum space (see, e.g., [43]). Here, $m$ is the mass of a molecule, i.e., twice the mass of a $^{87}$Rb atom. Figure 4 shows the calculated energy eigenvalues as a function of the lattice depth. The energies are given in terms of the recoil energy $E_R = h^2/(2m\lambda^2)$, with $h$ being Planck's constant. As we do not specify the quasimomentum, the energy eigenvalues form bands which are broad for low lattice depths. However, the bands $n = 0$ to $n = 2$ are quite narrow for lattice depths above $\sim 40 E_R$. This is the regime where we take most of our measurements. Having measured the resonant excitation frequencies after modulation we could in principle use Fig. 4 to read off the corresponding lattice depth $|U|$. We refine this method and at the same time check for consistency as follows.

In the experiment we control the lattice depth $|U|$ via the laser beam power $P$ that can be measured using photodiodes. The square to the electrical field $E_0^2$ is proportional to $P$. Consequently $|U| \propto P$, i.e., the precise value of $|U|$ is known up to a calibration factor (which depends linearly on the dynamical polarizability). Thus, given a molecular state, we should be able to adjust the calibration factor such that all data obtained for various powers $P$ match the band structure calculation. The measured data points in Fig. 4 clearly show that this works quite well, both for deeply bound molecules (red) and Feshbach molecules (blue). In this procedure we do not account for the transitions from $n = 0$ to $n = 4$ owing to the large uncertainties of the corresponding resonances in the excitation spectra. Nevertheless, these data points are shown in the plot for comparison.

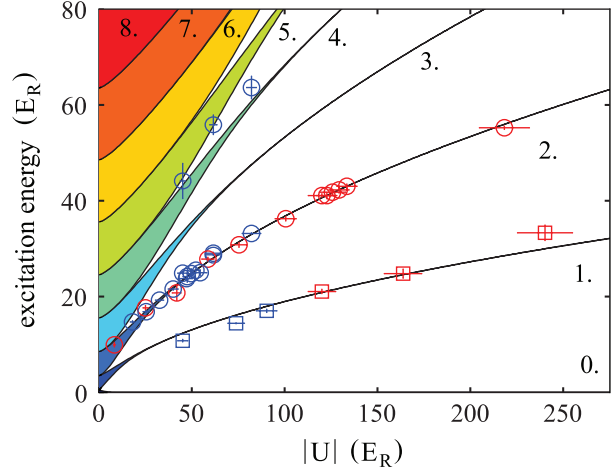

**Figure 4.** Energy band-structure with zero energy corresponding to the center of the lowest Bloch band. Solid lines are calculations for a single lattice direction, where the numbers 1. to 8. give the band index $n$. The data points are obtained excitation energies for $\lambda = 1064.5$ nm, stemming from amplitude (circles) or phase (squares) modulation spectroscopy. Red (blue) plot symbols indicate measurements for triplet rovibrational ground state (Feshbach) molecules. Here, the experimental results are shown after independently fitting the data for each molecular species to the band-structure calculation. By doing so, we determine the individual calibration factors and therefore the lattice depths $|U|$ (see also text). The horizontal error bars represent the resulting uncertainties of $|U|$, whereas the vertical error bars are given by the uncertainties of the Lorentzian fits in the modulation spectra.

In addition to $|U|$, the electrical field amplitude $E_0$ of the optical lattice has to be determined in order to infer the dynamical polarizability $\alpha(\omega)$ [see Eq. (1)]. We can circumvent this by referencing the measurements on the lattice depth $|U|$ for the molecules in the rovibrational ground state of the $a^3\Sigma_u^+$ potential to similar measurements with Feshbach molecules, of which the polarizability $\alpha_{\text{Fesh}}(\omega)$ is known to be twice the one of a Rb atom $\alpha_{\text{Rb}}(\omega)$ in the electronic ground state [19]. According to Eq. (1) the lattice depths $|U_{v_a=0}|$ and $|U_{\text{Fesh}}|$ for the $v_a = 0$ and Feshbach molecules are related by

$$\frac{|U_{v_a=0}|}{|U_{\text{Fesh}}|} = \frac{\text{Re}(\alpha_{v_a=0})}{\text{Re}(\alpha_{\text{Fesh}})} \tag{6}$$

for a given lattice beam intensity, i.e., a given $E_0$.

From our experiments at $\lambda = 1064.5$ nm we obtain $\text{Re}\{\alpha_{v_a=0}\} = (2.5 \pm 0.1) \times \text{Re}\{\alpha_{\text{Fesh}}\}$, whereas $\text{Re}\{\alpha_{v_a=0}\} = (0.1 \pm 0.02) \times \text{Re}\{\alpha_{\text{Fesh}}\}$ was found at $\lambda = 830.4$ nm [13]. Using our calculated atomic values of 685.8 a.u (1064.5 nm) and 2995.9 a.u (830.4 nm) yields molecular polarizabilities $\text{Re}\{\alpha_{v_a=0}\}$ of $3430 \pm 140$ a.u (1064.5 nm) and $600 \pm 120$ a.u (830.4 nm), respectively.

## 5. Comparison of results

Table 1 shows our measured and calculated polarizabilities along with results of other references. First, it should be noted that our theoretical atomic polarizabilities (including only the $5s - 5p$ and $5s - 6p$ transition frequencies from the NIST database [44] at 828.4 nm and at 1060.1 nm are in good agreement with the ones of Refs. [45, 46] which consider the $5s - 5p$ transition frequency from the NIST database, and *ab initio* values for the frequencies up to the $5s - 8p$ transitions.

We find good agreement between our theoretical (3147 a.u.) and experimental (3430 ± 140 a.u.) results for the molecular polarizability $\mathrm{Re}\{\alpha_{v_a=0}\}$ at 1064.5 nm (see table 1). In contrast, the agreement for the polarizability at 830.4 nm of our former measurements [13] (600 ± 120 a.u.) with the present calculations (875.8 a.u.) is somewhat poor. In view of this discrepancy we want to estimate the influence of slight shifts of the potential energy curves on the calculations. The potential well depths of the $1^3\Sigma_g^+$ and the $1^3\Pi_g$ states used in the computation are smaller by $72\,\mathrm{cm}^{-1}$ and $51\,\mathrm{cm}^{-1}$ with respect to the experimental determinations of Refs. [31] and [32], respectively. Such shifts would lead to a change in the calculated polarizability of about 10% for the particular wavelengths of 830.4 nm and 1064.5 nm. This sets a range for the uncertainty of the calculated polarizability that arises from the uncertainty of the PECs. Furthermore, in terms of the experiments, we note that in Ref. [13] a different method to determine the dynamical polarizability was used. The polarizability was inferred from the oscillating dynamics of molecular wave packets that occurred when $v = 0$ molecules were suddenly loaded into several Bloch bands of the optical lattice. This leaves potentially room for a systematic discrepancy between the two measurements. With respect to the static polarizability, i.e., $\omega = 0$, the calculations presented in this work give 698.5 a.u. (cf. table 1) for the $v_a = 0$ molecules. This value actually agrees well with the one previously reported in Ref. [51], 677.5 a.u., as it was not including the contribution of $\alpha_c \equiv 2 \times \alpha(\mathrm{Rb}^+) = 18.2\,\mathrm{a.u.}$ [47].

Figure 5 is a zoom into Fig. 1(b) showing the calculated real part of the dynamical polarizability of a $v_a = 0$ molecule (solid black lines). In addition, $\mathrm{Re}\{\alpha(\omega)\}$ for a Feshbach molecule is plotted (dashed red lines), which is given by twice the atomic polarizability. The two wavelengths used in our experiments (830.4 nm and 1064.5 nm) are indicated as vertical green dashed lines. Outside the resonant and therefore lossy regions in Fig. 5 (vertical black bands) the two polarizability curves never cross. Thus, there is no so-called "magic" wavelength, where the ac Stark shift of the two molecular states caused by the trapping light is equal. Such state-insensitive trapping conditions can be beneficial, e.g., when converting Feshbach molecules to deeply bound states. Specifically, in Ref. [13], owing to the large difference of the dynamical polarizabilities at $\lambda = 830.4$ nm, the STIRAP transfer of $\mathrm{Rb}_2$ from the Feshbach level to $v_a = 0$ populated several lattice bands.

We have again studied this issue in this work and find that population of higher lattice bands can be suppressed even in the absence of a magic wavelength when working with deep lattices. At $\lambda = 1064.5$ nm there is still a factor of 2.5 difference in polarizability between Feshbach and $v_a = 0$ molecules. For an initial (final) lattice depth of $50\,E_R$ ($125\,E_R$) at 1064.5 nm a calculation of the wavefunction overlap for the Bloch states shows that still 97% of the population stays in the lowest Bloch band after the STIRAP. In addition, we are able to energetically resolve the lattice bands during STIRAP as $n = 0$ and $n = 2$ are separated by about 40 kHz at $|U| = 125\, E_R$ (cf. Fig. 4). This strongly increases the selectivity of the transition (see, e.g., [52]). Indeed, in our experiments we do not observe any significant population of higher bands.

**Table 1.** Measured and calculated polarizabilities Re$\{\alpha\}$ in a.u. for $^{87}$Rb atoms and $^{87}$Rb$_2$ molecules in the rovibrational ground state of a$^3\Sigma_u^+$. Here, the abbreviations "tw exp (theo)" mean "this work, experimental (theoretical)".

| Species | $\lambda$ (nm) | Re$\{\alpha\}$ (a.u.) | Ref. |
|---|---|---|---|
| $^{87}$Rb | 828.4 | $3132 \pm 3$ | [45] |
| $^{87}$Rb | 828.4 | 3131.4 | tw theo |
| $^{87}$Rb | 830.4 | 2995.9 | tw theo |
| $^{87}$Rb$_2$ | 830.4 | 875.8 : | tw theo |
| $^{87}$Rb$_2$ | 830.4 | $600 \pm 120$ | [13], using $\alpha_{\text{Rb}} = 2995.9$ a.u. |
| $^{87}$Rb | 1060.1 | 692.7 | tw theo |
| $^{87}$Rb | 1060.1 | $693.5 \pm 0.9$ | [46] |
| $^{87}$Rb | 1064.5 | 685.8 | tw theo |
| $^{87}$Rb$_2$ | 1064.5 | 3147 | tw theo |
| $^{87}$Rb$_2$ | 1064.5 | $3430 \pm 140$ | tw exp, using $\alpha_{\text{Rb}} = 685.8$ a.u. |
| $^{87}$Rb$_2$ | 1064.5 | $3200 \pm 500$ | [48] |
| $^{87}$Rb | $\infty$ | $318.6 \pm 0.6$ | [50] |
| $^{87}$Rb | $\infty$ | 317.9 | tw theo |
| $^{87}$Rb$_2$ | $\infty$ | 698.5 | tw theo |
| $^{87}$Rb$_2$ | $\infty$ | 677.5 | [51] |

As can be seen in Fig. 5, the absolute dynamical polarizability of the triplet rovibrational ground state molecules at 1064.5 nm is about four times larger than at 830.4 nm. This is convenient since it results in a four times deeper interaction potential at the same laser intensity. For longer wavelengths than 1064.5 nm, Fig. 5 reveals, that the dynamical polarizabilities of $v_a = 0$ molecules and Feshbach molecules approach each other. Hence working at even longer wavelengths than 1064.5 nm might be advantageous for some applications.

## 6. Lifetime of the molecules

According to Eq. (2), the imaginary part of the dynamical polarizability, Im$\{\alpha\}$, is linked to the light power absorbed by a molecule, $P_{\text{abs}}$, which in turn can be expressed in terms of the photon scattering rate $\Gamma_{sc} = \hbar^{-1}\omega^{-1}P_{abs}$ [22]. Using this and Eq. (1), $\Gamma_{sc}$ can be written as

$$\Gamma_{sc} = -\frac{2U}{\hbar}\frac{\text{Im}\{\alpha(\omega)\}}{\text{Re}\{\alpha(\omega)\}}. \tag{7}$$

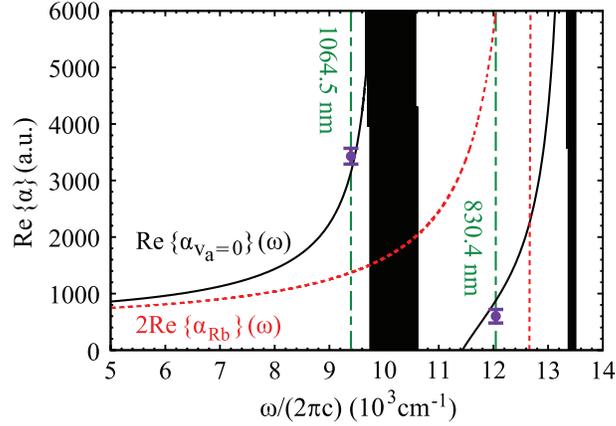

**Figure 5.** Real parts of the dynamical polarizabilities of triplet rovibrational ground state molecules (solid black lines) and Feshbach molecules (dashed red lines). The circles represent the experimental results given in table 1 with the corresponding wavelengths indicated by green dashed vertical lines.

For a 3D optical lattice with equal lattice depths $|U|$ in each direction the scattering rate is given by $\Gamma_{sc}^{3D} = 3\Gamma_{sc}$. As an example, we consider the case of $|U| = 50\,E_R$ at $\lambda = 1064.5$ nm. Then, the corresponding values for the polarizability obtained in the present work, $\text{Im}(\alpha) = 0.96 \times 10^{-3}$ a.u and $\text{Re}\{\alpha\} = 3430$ a.u., yield $\Gamma_{sc} = 0.18\,\text{s}^{-1}$. Note, this calculation only accounts for an ideal optical lattice. As there is always background light that does not contribute to the lattice, the estimated value for the scattering rate represents just a lower bound. Once a photon is absorbed, the molecule is excited and typically decays to a nonobservable state. Assuming excitation to be the only loss-mechanism a lifetime $\tau = 1.9\,\text{s}$ of the $v_a = 0$ molecules is expected in a $50\,E_R$ deep 3D optical lattice at $1064.5$ nm.

We experimentally investigate the lifetimes of the molecules in the rovibrational ground state of $a^3\Sigma_u^+$ by varying the holding time $t_h$ in the lattice. Figure 6 shows lifetime measurements of $v_a = 0$ molecules for various potential depths $|U|$, which are adjusted to be equal in each direction. Applying an exponential fit, we obtain a $1/e$ decay time $\tau$ of more than $2\,\text{s}$ for both our measurements at $|U| = 33\,E_R$ and $|U| = 58\,E_R$.

In order to estimate possible loss induced by inelastic molecular collisions, we calculate the tunneling rates $\Gamma_{tu}$ between adjacent lattice sites within the lowest Bloch band ($n = 0$). When considering a lattice depth of $33\,E_R$ ($58\,E_R$) one obtains $\Gamma_{tu} = 2.37\,\text{s}^{-1}$ ($\Gamma_{tu} = 0.09\,\text{s}^{-1}$). In our setup at most 20% of the lattice sites are occupied in the region of highest molecule density. Thus, for $|U| = 33\,E_R$ decay due to collisions cannot be neglected, whereas for $|U| = 58\,E_R$ and beyond the only relevant loss mechanism is photon scattering. For such deep lattices, the lifetime $\tau$ scales directly inversely with the lattice depth $|U|$. We confirm this for the measurements at $|U_1| = 58\,E_R$ and $|U_2| = 125\,E_R$, since the ratio of the lifetimes $\tau_1/\tau_2 = 2.20$ is close to $|U_2|/|U_1| = 2.16$.

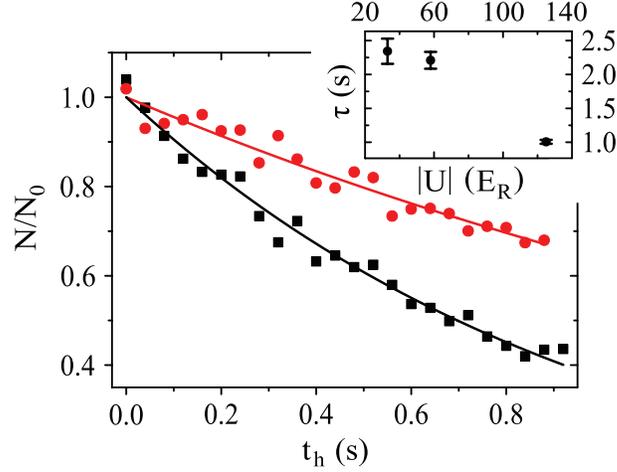

**Figure 6.** Decay of triplet rovibrational ground state molecules trapped in a 3D optical lattice at 1064.5 nm with equal potential depths $|U|$ in each direction. Shown is the fraction of remaining molecules $N/N_0$ as a function of the holding time $t_h$ in the lattice. Square plot symbols (red circles) correspond to a lattice depth of $125\,E_R\ (58\,E_R)$. Each data set typically consists of 10 to 15 repetitions of the experiment, where the statistical error of a data point is on the order of $\pm 0.1$. Solid lines are exponential fits to the data. Here, $N_0$ is given by the values of these fits at $t_h = 0\,\mathrm{s}$. In general, the absolute molecule numbers $N_0$ are about $1.5 \times 10^4$. The inset depicts the resulting $1/e$ decay times for various lattice depths.

## 7. Polarizability of $X^1\Sigma_g^+$ molecules

As the agreement between calculations and measurements for the triplet molecules is in general good, we also provide calculations for the singlet ground state of $^{87}\mathrm{Rb}_2$. Using the same approach as above (see also [19]), we compute the dynamical polarizability $\alpha_{v_X=0}(\omega)$, i.e., with respect to the $v_X = 0$, $J = N = 0$ level. The $X^1\Sigma_g^+$ PEC has been derived from spectroscopic data of Ref. [53]. The $A^1\Sigma_u^+$ and the $b^3\Pi_u$ PECs and the related spin-orbit coupling between those two states are taken from Ref. [54]. The PECs for all the other states and for the TEDMs are taken from the computations reported in Refs. [25, 26]. Again, the sum in Eq. (3) has been truncated to include only the levels of the four lowest $^1\Sigma_u^+$ states and the three lowest $^1\Pi_u$ states. The natural lifetime of the excited levels has been fixed at 10 ns. Results are presented in Fig. 7, showing that two magic wavelengths can be identified at 990.1 nm and 1047.2 nm. The latter is located close to the region of strong absorption resonances and consequently, from the imaginary part [cf. Fig. 7(b)], the photon scattering rate at 1047.2 nm is expected to be about four times larger than at 990.1 nm.

## 8. Parametrization of the polarizability

### 8.1. Rovibrational ground state of $a^3\Sigma_u^+$

In general, using figures (e.g., Figs. 1 and 7) to read off the dynamical polarizabilities at specific wavelengths is cumbersome. Therefore, we provide here a simple analytical fitfunction and parameters that allow for reproducing the numerical results with respect to nonresonant wavelength regimes. In the infrared and optical domain we are studying, the main contributions to the polarizability of the X (a) state outside of resonances come from the transitions towards the first excited $^{1(3)}\Sigma^+$ state and the first $^{1(3)}\Pi$ state. Thus we attempt to model the polarizability by reducing those transitions to a single effective transition towards each of the two different symmetries. The approximate real part of the polarizability is expressed as

$$\alpha_{\text{eff}}(\omega) = \frac{2\omega_\Sigma}{\hbar(\omega_\Sigma^2 - \omega^2)}d_\Sigma^2 + \frac{2\omega_\Pi}{\hbar(\omega_\Pi^2 - \omega^2)}d_\Pi^2 + \alpha_c \tag{8}$$

with $\omega_\Sigma$ (resp. $\omega_\Pi$) the effective transition frequencies and $d_\Sigma$ (resp. $d_\Pi$) the corresponding effective dipole moments. We have isolated in this expression the core polarizability $\alpha_c$ as its frequency dependence is much weaker than the one of the terms coming from valence electron excitation (see Appendix A.1). The imaginary part of the polarizability is neglected since the model is designed for the ranges outside the resonant regions.

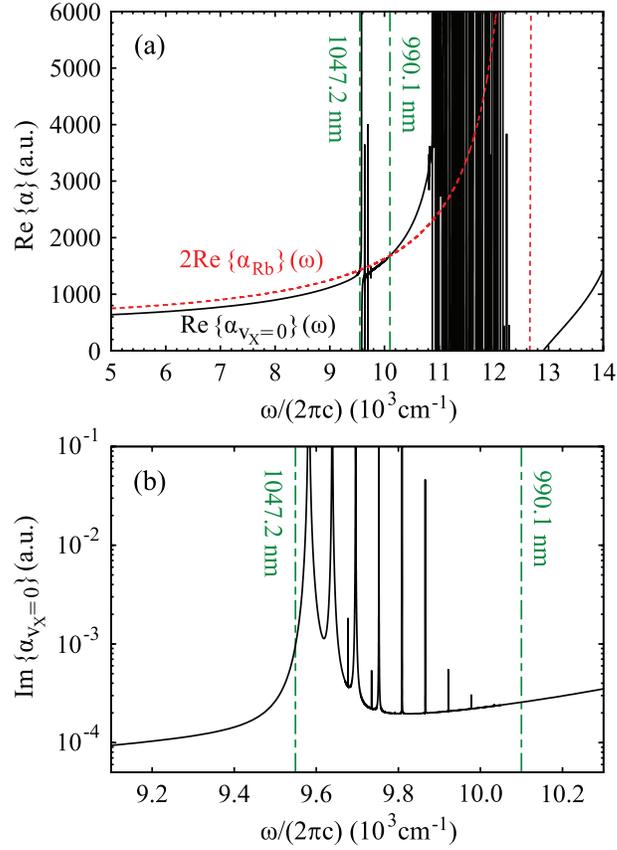

**Figure 7.** Real part (a) and imaginary part (b) of the dynamical polarizability of $^{87}$Rb$_2$ X$^1\Sigma_g^+$ molecules in their $v_X = 0$, $N = 0$ level. In (a), for comparison, also the numerical results corresponding to $2\text{Re}\{\alpha_{Rb}\}$ are shown (red dashed line), representing the real part of the dynamical polarizability of Feshbach molecules. Two magic wavelengths, where both polarizabilities are equal, are obtained and indicated by vertical dashed lines.

We extract the effective parameters from a fit to the full numerical results using Eq. (8). The results with respect to the a$^3\Sigma_u^+$ and X$^1\Sigma_g^+$ rovibrational ground state are shown in Fig. 8. For the triplet case, data points with frequencies close to resonances, i.e., from $\omega/(2\pi c) = 9527\,\text{cm}^{-1}$ to $11018\,\text{cm}^{-1}$ and above $13173\,\text{cm}^{-1}$, are excluded from the fit. We obtain $\omega_\Sigma/(2\pi c) = 10112.33\,\text{cm}^{-1}$, $d_\Sigma = 2.792881\,\text{a.u.}$, $\omega_\Pi/(2\pi c) = 13481.32\,\text{cm}^{-1}$ and $d_\Pi = 3.211023\,\text{a.u.}$ Note, in terms of dipole moments, atomic units can be converted into SI units according to $1\,\text{a.u.} = ea_0 = 8.478 \times 10^{-30}\,\text{Ams}$. Using these parameters, the effective polarizability reproduces the numerical results in the fitted region to within a relative root mean square value (rRMS) of around 1% [see Fig. 8(a)]. The rRMS value is defined by

$$\text{rRMS} = \sqrt{\sum_{i=1}^{M} \frac{1}{M} \left( \frac{\alpha_\text{eff}(\omega_i) - \alpha(\omega_i)}{\alpha(\omega_i)} \right)^2} \qquad (9)$$

with $\alpha(\omega_i)$ being the numerical values, and $\alpha_\text{eff}(\omega_i)$ the fitted ones. Here, $M$ is the number of considered values $\omega_i$, which depends on the vibrational level as the resonant frequency regions exluded from the fit vary. We point out that if we take the transition dipole moment $d_Z$ (resp. $d_X = d_Y$) at the equilibrium distance of the $a^3\Sigma_u^+$ state and multiply it by the appropriate Hönl-London factor ($1/\sqrt{3}$ for $\Sigma^+ - \Sigma^+$ transition and $\sqrt{2/3}$ for $\Sigma^+ - \Pi$ transition), we get the value $2.767\,\text{a.u.}$ (resp. $3.142\,\text{a.u.}$), very close to the effective dipole moment found above. Moreover, the effective transition frequencies correspond roughly to the average frequencies of transitions with favorable Franck-Condon factor.

### 8.2. Rovibrational ground state of $X^1\Sigma_g^+$

A similar fit can be performed for the dynamical polarizability of $X^1\Sigma_g^+$ molecules in the ($v_X = 0$, $N = 0$) level [Fig. 8(b)]. Frequency domains from $\omega/(2\pi c) = 9432\,\text{cm}^{-1}$ to $9960\,\text{cm}^{-1}$, from $10613\,\text{cm}^{-1}$ to $12890\,\text{cm}^{-1}$ and above $14736\,\text{cm}^{-1}$, corresponding to resonances towards the $b^3\Pi_u$, $A^1\Sigma_u^+$ and $B^1\Pi_u$ excited states, respectively, were excluded from the fit. We omitted to account for levels of the $b^3\Pi_u$ state as they have a very small singlet character. Then, the fit to the numerically calculated dynamical polarizability yields $\omega_\Sigma/(2\pi c) = 11450.31\,\text{cm}^{-1}$, $d_\Sigma = 2.647731\,\text{a.u.}$, $\omega_\Pi/(2\pi c) = 15019.57\,\text{cm}^{-1}$ and $d_\Pi = 2.965774\,\text{a.u.}$ Again, we point out that the effective transition dipole moments obtained are very close to the electronic transition dipole moments taken at equilibrium distance multiplied by the Hönl-London factors, i.e., $2.613\,\text{a.u.}$ and $2.959\,\text{a.u.}$, respectively. Here, the rRMS of the fit is 0.5%.

### 8.3. Excited vibrational states

Both, the frequency domains with good Franck-Condon factors and transition dipole moments depend significantly on the initial vibrational level. This is reflected in the variation of the effective parameters when we perform an individual fit for each vibrational level (with $N = 0$) of the $a^3\Sigma_u^+$ ($v_a = 0$ to 40) and $X^1\Sigma_g^+$ ($v_X = 0$ to 124) states. The corresponding parameters, which can be used to reproduce the dynamical polarizabilities in the nonresonant frequency domains, are reported in the Supplemental Material [20]. Figure 9 shows the resulting rRMS values as a function of the vibrational level with respect to the lowest triplet and singlet state.

Using the ansatz of Eq. (8) gives a poor result for most excited vibrational levels of the $a^3\Sigma_u^+$ potential with an rRMS exceeding 4% in the range of $v_a = 2$ to 27. This behavior is related to the fact that Franck-Condon factors mainly depend on the amplitude of the excited wave function around both the inner and the outer classical turning points of the initial vibrational level studied. The shape of the $a^3\Sigma_u^+$ potential is quite different compared to the relevant excited states. Consequently, the inner turning point region of a given excited vibrational level $v_a$ will induce couplings to levels of excited molecular states with energies strongly different from those related to couplings induced by the outer turning point region. This effect mainly occurs for the excited $^3\Sigma_g^+$ potential wells as they are deep. Instead, the depth of the $^3\Pi_g$ potential is not sufficient to create such a variation.

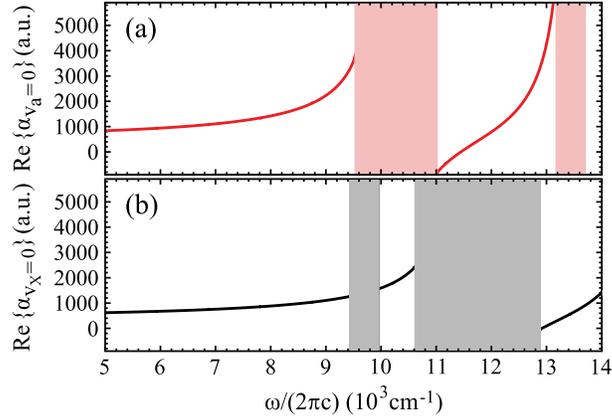

**Figure 8.** Analytical fits of the dynamical polarizabilities of the $a^3\Sigma_u^+$ molecule in $v_a = 0$, $N = 0$ (a), and of the $X^1\Sigma_g^+$ molecule in $v_X = 0$, $N = 0$ (b). Frequency ranges excluded from the fits are indicated as shaded areas. At the scale of the figure, the original curves and the fitted curves are indistinguishable.

Thus we added one more transition term to the ansatz of Eq. (8) in order to account for both the inner and outer part of the $a^3\Sigma_u^+$-$^3\Sigma_g^+$ transitions in the model. This reduced significantly the rRMS of the effective polarizabilities of the $v_a$ levels (see Fig. 9). We want to emphasize that such an interpretation gives a reasonable physical picture for most levels. However, for some levels like the deeply bound ones or those close to the dissociation limits, the three-effective-transition model is somewhat artificial and the effective parameters should be taken only as numerical parameters needed to easily obtain the corresponding polarizability.

## 9. Conclusion

We have studied the dynamical polarizability of the $^{87}$Rb$_2$ molecule in the rovibrational ground state of the $a^3\Sigma_u^+$ potential. Calculations of both, the real and imaginary part are provided and we measured Re$\{\alpha(\omega)\}$ at $\lambda = 1064.5$ nm. Our experimental and theoretical findings show good agreement. From our computed value of Im$\{\alpha(\omega)\}$ at this wavelength, we expect trapping times of the molecules on the order of seconds for lattice depths around $50\,E_R$, which was confirmed by our observations. We also have investigated theoretically the dynamical polarizability of the singlet ground state $X^1\Sigma_g^+$. These results are interesting for future STIRAP transfer of Rb$_2$ to the corresponding rovibronic ground state. Furthermore, we have introduced a simple analytical expression to parametrize the dynamical polarizabilities for all levels of both, $a^3\Sigma_u^+$ and $X^1\Sigma_g^+$ states. By fitting this expression to the numerical results, we have extracted effective parameters, which can be used to reproduce $\mathrm{Re}\{\alpha(\omega)\}$ of a given vibrational state with high fidelity. The precise knowledge of the dynamical polarizability enables accurate control of optical dipole potentials and therefore is of importance for future experiments with deeply bound $\mathrm{Rb}_2$ molecules.

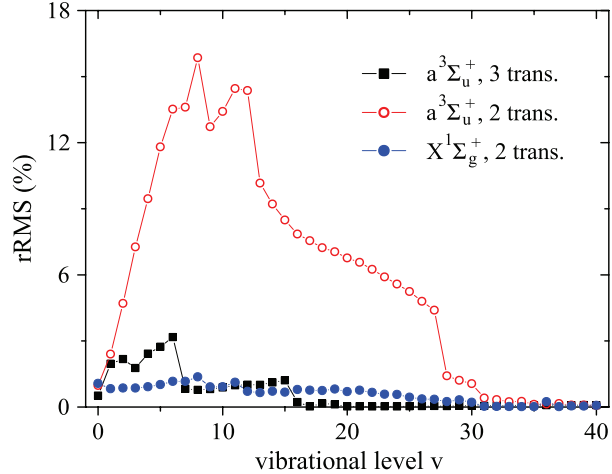

**Figure 9.** Relative root mean square (rRMS) values of the analytical fits to the numerically calculated dynamical polarizabilities. Blue closed (red open) circles correspond to fit results obtained with the "two-effective-transitions" ansatz for the vibrational levels $v_a$ of $a^3\Sigma_u^+$ ($v_X$ of $X^1\Sigma_g^+$), whereas the black closed squares stem from fits including a third effective transition to represent the dynamical polarizabilities with respect to $v_a$, which significantly increases the accuracy (see text). The rRMS values for $v_X > 40$ do not exceed 0.1% (not shown here).

## Acknowledgments

This work was funded by the German Research Foundation (DFG). R.V. acknowledges partial support from Agence Nationale de la Recherche (ANR), under the project COPOMOL (contract ANR-13-IS04-0004-01).

## Appendix

### A.1. Polarizability of the $Rb^+$ core

As the two ionic $\mathrm{Rb}^+$ cores are only weakly perturbing each other, we consider the molecular core polarizability as twice the atomic $\mathrm{Rb}^+$ polarizability. First, we calculate the atomic polarizabilities at imaginary frequencies $\alpha_{\mathrm{Rb}}(i\omega)$ [55], which includes only the contribution of the valence electron. We subtract them from the values of Ref. [23], where the resulting differences represent the contribution of the core electrons, and thus the $\mathrm{Rb}^+$ polarizability. Following Ref. [23] this estimate assumes that the influence of the valence electrons on the core polarizability is negligible.

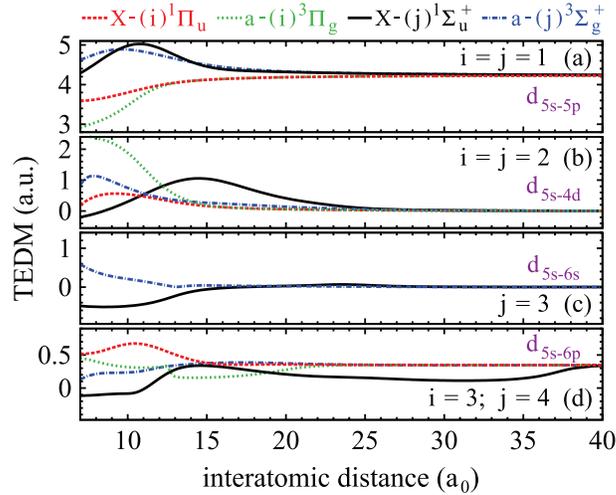

**Figure 10.** Computed transition electric dipole moments (TEDMs) for the main transitions from the $a^3\Sigma_u^+$ and the $X^1\Sigma_g^+$ states to the states correlated to the $5s + 5p$ dissociation limit (a), the $5s + 4d$ dissociation limit (b), the $5s + 6s$ dissociation limit (c), and the $5s + 6p$ dissociation limit (d). The individual curves can be assigned to the corresponding transitions as indicated on top of the graph using the numbers $i$ and $j$. At large distances the TEDMs converge towards $d_{5s-nl}$ which is equal to the atomic TEDMs multiplied by $\sqrt{2}$ (see text).

We use an ansatz similar to the one of Eq. (8) to model the core polarizability with two effective transitions. This yields the effective frequencies $\omega/(2\pi c) = 165912\,\mathrm{cm}^{-1}$ and $362918\,\mathrm{cm}^{-1}$, and the effective dipole moments $2.72799$ a.u. and $1.40119$ a.u. We note that these two transition frequencies are on the order of magnitude of the main transition in $\mathrm{Rb}^+$ and $\mathrm{Rb}^{2+}$, respectively. The core polarizability is a small contribution slowly varying from $17.9$ a.u. at vanishing frequency up to $18.1$ a.u. in the optical domain relevant here. For the static case, we can compare the result of our calculations to the value of $18.2$ a.u. reported in Ref. [47] and find good agreement.

### A.2. Transition electric dipole moments

For the sake of completeness we show in figure 10 the transition electric dipole moments included in the calculations of the dynamical polarizabilities as functions of the internuclear distances. The corresponding numerical values are given in the Supplemental Material [20]. Some of the data with respect to the lowest transitions have already been reported in Refs. [25, 26]. The largest TEDMs in the range of the PECs concern the first excited $\Sigma^+$ and $\Pi$ potentials [see Fig. 10(a)]. At large distances the molecular excited electronic wave functions become close to the form $[\phi_{5s}(1)\phi_{nl}(2) \pm \phi_{5s}(2)\phi_{nl}(1)]/\sqrt{2}$ and therefore the TEDMs converge towards $d_{5s-nl}$, which is equal to the atomic TEDMs multiplied by $\sqrt{2}$. We find $d_{5s-5p} = 4.23$ a.u. and $d_{5s-6p} = 0.347$ a.u., in excellent agreement with the values extracted from the NIST database [44] ($4.226$ a.u. and $0.3531$ a.u.), obtained by averaging the TEDMs corresponding to $5s - np_{1/2}$ and $5s - np_{3/2}$ for $n = 5$ and 6, respectively. This confirms the good quality of the present representations of the atomic electronic wave functions.